\documentclass{article}[12pt, a4paper]

\usepackage[english]{babel}
\usepackage[dvipsnames]{xcolor}
\usepackage[a4paper,top=2cm,bottom=2cm,left=3cm,right=3cm,marginparwidth=1.75cm]{geometry}
\usepackage{authblk}
\AtBeginDocument{%
  \providecommand\BibTeX{{%
    \normalfont B\kern-0.5em{\scshape i\kern-0.25em b}\kern-0.8em\TeX}}}

\usepackage{amsmath}
\usepackage{graphicx}
\usepackage[colorlinks=true, allcolors=blue]{hyperref}
\usepackage{cite}
\usepackage{soul}
\usepackage{array}
\usepackage{multirow}
\usepackage{subcaption}
\usepackage{tabularx}
\usepackage{booktabs}
\usepackage{float}

\graphicspath{{Figures/}}
\newcommand{\PreserveBackslash}[1]{\let\temp=\\#1\let\\=\temp}
\newcolumntype{C}[1]{>{\PreserveBackslash\centering}p{#1}}

\title{A process mining-based error correction approach to improve data quality of an IoT-sourced event log}

\author[1]{Mohsen Shirali}
\author[2]{Zahra Ahmadi}
\author[3,4]{Carlos Fernández-Llatas}
\author[3]{Jose-Luis Bayo-Monton}
\author[5]{Gemma Di Federico}

\affil[1]{Computer Science and Engineering, Shahid Beheshti University, Tehran 19839-63113, Iran (\texttt{m\_shirali@sbu.ac.ir})}
\affil[2]{Research Centre for Information Systems Engineering (LIRIS), KU Leuven, Brussels 1000, Belgium (\texttt{zahra.ahmadi@kuleuven.be})}
\affil[3]{Process Mining 4 Health Lab–SABIEN-ITACA Institute, Universitat Politècnica de València. Valencia 46022, Spain (\texttt{jobamon@itaca.upv.es, cfllatas@itaca.upv.es})}
\affil[4]{Department of Clinical Sciences Intervention and Technology (CLINTEC), Karolinska Institutet, Stockholm 17177, Sweden}
\affil[5]{Technical University of Denmark, Kgs. Lyngby, 2800, Denmark(\texttt{gdfe@dtu.dk}}
\date{}

\begin{document}

\maketitle
{\small{Corresponding author: Mohsen Shirali (\texttt{m\_shirali@sbu.ac.ir})}}

\begin{abstract}
Internet of Things (IoT) systems are vulnerable to data collection errors and these errors can significantly degrade the quality of collected data, impact data analysis and lead to inaccurate or distorted results.
This article emphasizes the importance of evaluating data quality and errors before proceeding with analysis and considering the effectiveness of error correction methods for a smart home use case.

\end{abstract}


\section{Introduction}
\label{1-intro}

The Internet of Things (IoT) has become a part of people's everyday lives and as smart devices become ubiquitous, they will continue to integrate into our day-to-day activities~\cite{pohls2014rerum}.
From a logical viewpoint, an IoT system can be depicted as a collection of low-cost smart sensors that can continuously collect different types of data, from ambient environments such as room temperature, light or humidity to personal data such as basic physiological data~\cite{zhao2020privacy}.
These smart sensors could be embedded within the devices people wear (like wristbands, smartwatches or smart glasses), the devices are used in a home or installed in an environment (including security cameras, smart assistants and smart locks or even simpler devices like smart thermostats and movement detectors)~\cite{apthorpe2017smart}. 

The IoT technology has the potential to create a new “cyber-physical” world, in which “things” can directly operate, act and influence the physical world.
Hence, the continuous flow of data from the physical to the digital world, provided by IoT devices, will extend the situational awareness of computers, thus, gaining the ability to act on behalf of humans through ubiquitous services~\cite{karkouch2016data}.
Indeed, the data collected by IoT devices, when mined using data mining techniques and algorithms, gives insights about a given phenomenon, person or entity which can be used to provide intelligent services~\cite{chun2013yang, sicari2015security}.
Thus, data is the main valuable asset of every IoT system and its quality is of great importance.
Using data with degraded quality, due to the occurrence of errors\footnote{The definition of error in this article is mainly close to the definition provided in articles like~\cite{teh2020sensorquality}, and we are not considering other definitions related to data quality in database-level or high-level applications.}, may complicate further analysis and lead to misleading results or wrong decisions~\cite{teh2020sensorquality, bertrand2022dataquality}.

Incomplete data and erroneous (or incorrect) data are two main subtypes of errors that mostly occur during data collection by IoT devices.
Both of these types of errors can compromise the quality of the collected data.
These errors can be caused by various factors, such as damaged or depleted sensor batteries, data loss or corruption at the network level, leading to unstable or incomplete sensor readings~\cite{branch2009network, zeng2011web, karkouch2016data}.
Additionally, environmental factors such as local weather conditions, improper device placement, range limitations, and device malfunctions can all affect sensor operation and result in incorrect data~\cite{teh2020sensorquality}.
In these cases, high-quality sensors such as industry-grade sensors could help, however, due to the need for deployment of a large number of sensors in a network, which is the case for many IoT applications, it is not always feasible to use expensive industry-grade sensors and most IoT applications use low-cost sensors, at the expense of losing data quality.
Furthermore, sometimes retransmitting packets to make up for lost data caused by wireless communication failure is not feasible. Reasons like the complexity of identifying lost packets for retransmission, the impracticality of deferring decision-making in delay-sensitive IoT applications, and power limitations prevent retransmission~\cite{li2014nearest,teh2020sensorquality}.

Moreover, when sensors' raw data is collected, they should be prepared before starting any analysis.
Typical sensor data preparation~\cite{Zhang2003DataPreparation} or pre-processing steps are cleaning, formatting and event abstraction to derive a meaningful event log from sensor data.
In this way, various types of errors in the sensor log, mainly stemming from the intrinsic characteristics of IoT sensors, could be propagated or amplified into the event log by event abstraction.
Therefore, these event log quality issues (e.g., incorrect events) are mostly inevitable and they greatly hamper the mining techniques~\cite{bertrand2022dataquality}.

Many studies express that sensor data quality plays a vital role in IoT applications and state the importance of data quality for data mining purposes~\cite{karkouch2016data, berti2007measuring, hand2001principles}.
The potential degraded quality issues stress the need for through data quality assessment, \textit{i.e.} determining whether data quality issues are present in the event log. Awareness regarding the prevailing data quality issues can help to take initiatives to alleviate them~\cite{Martin2021DataQuality}. 
In this way, to take the most out of IoT technology's full potential, before starting any analysis and applying mining techniques on data, its quality and accuracy should be ensured to prevent misguided decisions~\cite{bertrand2022dataquality}. If it is possible, the errors should be detected or quantified and removed or corrected in order to improve sensor data quality~\cite{teh2020sensorquality}.

Several methods have been proposed to detect, remove or correct the errors before applying data mining algorithms.
However, the challenge is that the existing methods, proposed to solve errors in sensor data, cannot be compared due to different evaluation processes.
In addition, the relation between the sources of errors in the log, their impact on deviating analyses and how the correction techniques can be designed and adjusted to error sources in order to properly correct the errors are not clear.
Therefore, this motivates us to investigate the impact of errors within a log collected in a real-life scenario and try to improve the log quality by correcting the errors.

To be more specific, we have focused on IoT usage in healthcare applications, which also suffer from data quality issues.
Despite the great potential of mining algorithms to analyse healthcare data, the reliability of their outcomes ultimately depends on the quality of the input data used by the algorithms.
Applying mining techniques to low-quality data can lead to counter-intuitive or even misleading results~\cite{andrews2018towards}.
For our study, we have chosen a smart home case study in healthcare and used a dataset collected by ambient binary sensors in a solo-resident house.
This particular scenario was selected to provide a consistent platform for comparing the effectiveness of multiple error correction techniques in addressing various types of errors.
We aim to investigate the impact of adjustments made to existing methods in addressing errors and examine our proposed hybrid approach for inspecting a log to identify possible errors.

We initially processed the event log to identify the error types that existed in the dataset and the most problematic sensors and events and employed a rule-based approach tailored to our case study.

The rest of this article is organized as follows:
\nameref{Section2:backgrounds} section provides the required information to familiarise readers with the AAL systems by looking into their applications, declares the definition of error, its types and the importance of addressing errors, and reviews the existing error correction methods in the literature.
In addition, the dataset, which is used in this research and its features are presented in this section.
Next,~\nameref{proposed_techniques} section describes the error correction techniques and how they are adapted to our use case for addressing errors.
Then, in the~\nameref{Section4:results-discussion} section, the results and performance of error correction techniques are evaluated. 
Finally, section~\nameref{Section5:Conclusions} finalise the article and sums up the article's findings.


\section{Background knowledge}
\label{Section2:backgrounds}

To have a common understanding of IoT systems and their applications in the healthcare domain, and Ambient Assisted Living (AAL) systems as the selected case study in this manuscript, we first introduce these systems in Section~\ref{Section2.1-AAL} to familiarise the readers with these systems and existing sensors for data collection.
Then, in Section~\ref{Section2.2-error} the error definition and its various types, which are applicable to the selected AAL scenario, are briefly defined to avoid any confusion and to establish a common ground for later usage.
Section~\ref{Section2.3-ErrorRW} shortly overviews the existing related approaches for error correction.
Finally, in Section~\ref{Section2.4-dataset} the details of the used dataset in this research, which is used for the evaluation of our proposed approach, is given.

\subsection{Ambient Assisted Living Systems}
\label{Section2.1-AAL}

AAL systems use various sensors to measure physical quantities such as temperature, humidity, and motion. These measurements, along with any changes captured by the sensors, are recorded in \textbf{data logs} (also known as \textbf{sensor logs}). An \textbf{event log}, on the other hand, records specific occurrences, activities or incidents that happen in the under-control environment. While events are happening, multiple entries are added to the data log by the sensors to capture any changes in their measurements.
One or more sensor measurements can then be converted into different events in the event log, which can be inferred from the raw sensor data using techniques like abstraction or aggregation.
Indeed, an event log provides a timeline of actions or incidents taken by users, systems, or devices, with every single event occurring at a given point in time. 
While sensor logs focus on recording raw data, event logs focus on significant events, incidents, and activities extracted or abstracted from the sensor raw readings in the sensor log.

\subsection{Error definition and types}
\label{Section2.2-error}
The term error refers to the discrepancy or difference between the measured or observed value of a particular data point and the true or expected value of that data point.
Different terms related to the error (sometimes not uniformly) have been used in the literature to describe its various types such as outliers, missing data, bias, drift, noise, constant value, etc.~\cite{teh2020sensorquality}.
However, in this section, we focus on the errors that can occur and exist in the log collected by AAL systems equipped with ambient binary sensors, as our target scenario (interested readers who want to study other error types, in more detail, can refer to other studies such as~\cite{teh2020sensorquality, Martin2021DataQuality}).

The binary sensors including PIRs and contact switches, can only provide two values: \textit{zero} and \textit{one}.
Thus, in a smart home with binary sensors, only specific kinds of errors can occur during data collection and decrease the quality of the sensor log.
We categorise the possible errors for an AAL system with binary sensors into two different types:

\begin{itemize}
    \item \textbf{Type 1) Missing events (incomplete data).}
    These are events that occurred in reality but were not recorded in the sensor/event log~\cite{bose2013wanna}.
    Indeed, the under-control environment is changed in such a way that sensors should be triggered, but sensors do not capture the actual event, or the event is captured but the transmission failed, hence the event is not added to the log.
    According to~\cite{li2014nearest}, unstable wireless connection due to network congestion or environmental interference \textit{e.g.} human blockage, walls, and weather conditions, sensor device outages due to its limited battery life and malicious attacks are the most frequently mentioned reasons for these errors.
    In addition, the sensor coverage (sensing area or field of view) is limited and the sensors can only detect changes in a specific zone.
    Thus, if sensors are not properly installed, their coverage might not encompass the whole area where the activities could take place.
    This issue leads to missing events that happened beyond the sensor’s reach. The sampling rate of sensors is limited as well and inadequate sampling rates can cause missing events in the event logs~\cite{Shirali2020Mobility, berti2007measuring}.
    
    \item \textbf{Type 2) Noises (erroneous events).}
    Event entries which are present, but for which the recorded values do not reflect reality~\cite{Martin2021DataQuality}.
    The noises happen when the sensors do not behave as expected and are present in the sensor data due to issues during logging (\textit{e.g.} transmission of packets and writing the reported data in the dataset) or due to the presence of conditions affecting the phenomenon measured by the sensors~\cite{bertrand2022dataquality}. Incorrect or inexactness of events' timing is one of the main subcategories of these errors.
    When there is a discrepancy between the moment at which an event took place in reality and the moment at which it is recorded in the system, the incorrect timing error occurs~\cite{vanbrabant2019quality}.
    Another reason for incorrect events is the noise in localization systems which can caused by unpredictable reflections, refractions, or absorptions of signals due to the heterogeneity of walls and furniture of rooms, environmental factors, or even humans, depending on their positions~\cite{Fernandez-Llatas2021RTLS}.
    In addition, an inaccurate sensor placement can also lead to erroneous events.
    When multiple sensors, which are installed to cover different areas, cover the same area or phenomenon (\textit{e.g.} the sensing area of multiple sensors overlap with each other), they create multiple records for a single event -occurred inside the overlapping area- with same timestamps and different labels, so that one of them is correct, while the others are noises.
    For instance, movement detection sensors installed in adjacent locations will detect a single movement simultaneously and generate multiple erroneous events with different labels in the log~\cite{Shirali2020Mobility}.
    
\end{itemize}

The other subtypes of errors like drifts that refer to changes in the reported values by sensors (\textit{e.g.} constant changes in the measured value, for instance, always reporting a value higher than the actual value) are not applicable to binary sensors.
In addition, some sub-types are impossible, undiscoverable or at some points equivalent to the considered types.
For instance, drifted values are impossible because we only have two outputs and reporting a constant value is equivalent to either undetected events, when the reported value is zero, or noisy detection if the value equals one.

The missing event errors can sometimes be addressed by performing appropriate actions, like re-transmission of sensor readings to compensate for the loss of packets containing data for captured events and accurate installation of sensors to avoid undiscoverable areas.
A post-evaluation of the log quality to find possible error sources and adjust the sensors' field-of-view and sampling rates is another effective solution to prevent missing event-type of errors.
On the other side, using more accurate sensors, modifying the placement and coverage of sensors to remove overlapping areas, and filtering out events that are obviously abnormal (for instance too short or too long) are possible solutions to handle noises.

However, modifying the configuration of the IoT system for data collection to improve the quality is not always possible and even high-accuracy sensors might have noises.
In addition, in some cases, data mining techniques are used to analyse offline datasets, which are collected before and cannot be changed.
Thus, it is mandatory to have solutions that can address errors in the sensor and event log after data collection.
Error detection and correction methods can be used in these situations to improve the quality of the data.

\subsection{Existing methods for error detection and correction}
\label{Section2.3-ErrorRW}
Several solutions have been proposed to quantify or detect errors in the literature.
In a study in \cite{teh2020sensorquality} different types of sensor data errors and solutions for detecting and correcting them were systematically reviewed, irrespective of the IoT architecture layer.

Data mining practitioners often overlook infrequent events to avoid incurring additional costs associated with error correction~\cite{dogan2019individual}.
These infrequent events are seen as noise and removed to reduce variability and limit the spaghetti effect.
On the other hand, there are also de-noising methods (such as~\cite{yang2017data}) that attempt to remove the noise associated with the measured value.
There are also missing data imputation methods that are used to estimate the missing sensor measurement values or the corresponding events.
Two such methods are clustering and k-nearest neighbour, which are used to estimate missing sensor values. 
 
In~\cite{Shirali2020Mobility}, a process mining technique is used to create a human-readable model for transitions between different areas of a house equipped with a PIR-based movement detection system.
The discovered model highlighted many invalid transitions caused by detection errors, enabling the system designers to identify the most problematic sensors.
By redeploying noisy sensors, the quality of the logged data was significantly improved.

In addition to improving the quality of logs, error correcting techniques can also be utilised in certain analysis applications.
\cite{chiloiro4713049impact} have proposed a method that uses these algorithms to provide a more comprehensive overview of compliance and identify the most common types of deviations.
They provide a conformance-checking approach that employs error-correcting methods to measure the degree of deviation from clinical guidelines in the treatment process of patients, by integrating healthcare professionals' knowledge.

\subsection{The case study dataset}
\label{Section2.4-dataset}
This study uses a dataset~\cite{falah2022probabilistic} collected from a house with a 60-year-old female resident.
It contains data from ambient sensors installed within a house, wristband data, smartphone information on the usage of mobile applications and daily psychological reports for a period of 147 days.
However, we only use data captured by ambient sensors in this study. 

Ambient sensorial information includes multiple PIR sensors showing the presence of the subject in different areas of the house. A power usage sensor indicating TV usage, contact sensors highlighting the opening and closing of the Bathroom, WC and closet doors, and a gas detection sensor that detects cooking activity.  
The ambient data is pre-processed to extract information about the subject’s presence in different areas of the house. Through this process, we have been able to detect events related to the visited home locations for the entire dataset period.

We evaluate this dataset, which we call it FR-dataset hereafter, in terms of errors.
In addition, we applied multiple approaches including a process mining-based error correction, a rule-based error filtering method and a combination of both on the location event log to investigate how these algorithms can handle the error problems.

\subsubsection{Locations and duration of events}
To determine the exact location of the resident inside the home using ambient sensorial information, we can look at the locations in which sensors are mounted and use the timestamp of logged data to infer the person’s location at different times~\cite{Shirali2020Mobility, wren2006toward, cook2009collecting}.
For instance, a record captured by the bedroom’s closet sensor indicates that the subject was in the bedroom at that time.
Then, by correlating each sensor to its installation location, every captured record can be mapped to a location.
Therefore, if we have the whole sequence of captured records for a specific duration, then events related to the person’s presence in different areas of the house and even the sequence of they movements can be inferred~\cite{Lull2021behaviour}.

\begin{figure}[t]
    \centering
    \begin{subfigure}[h]{0.6\textwidth}
        \centering
        \includegraphics[width=\textwidth]{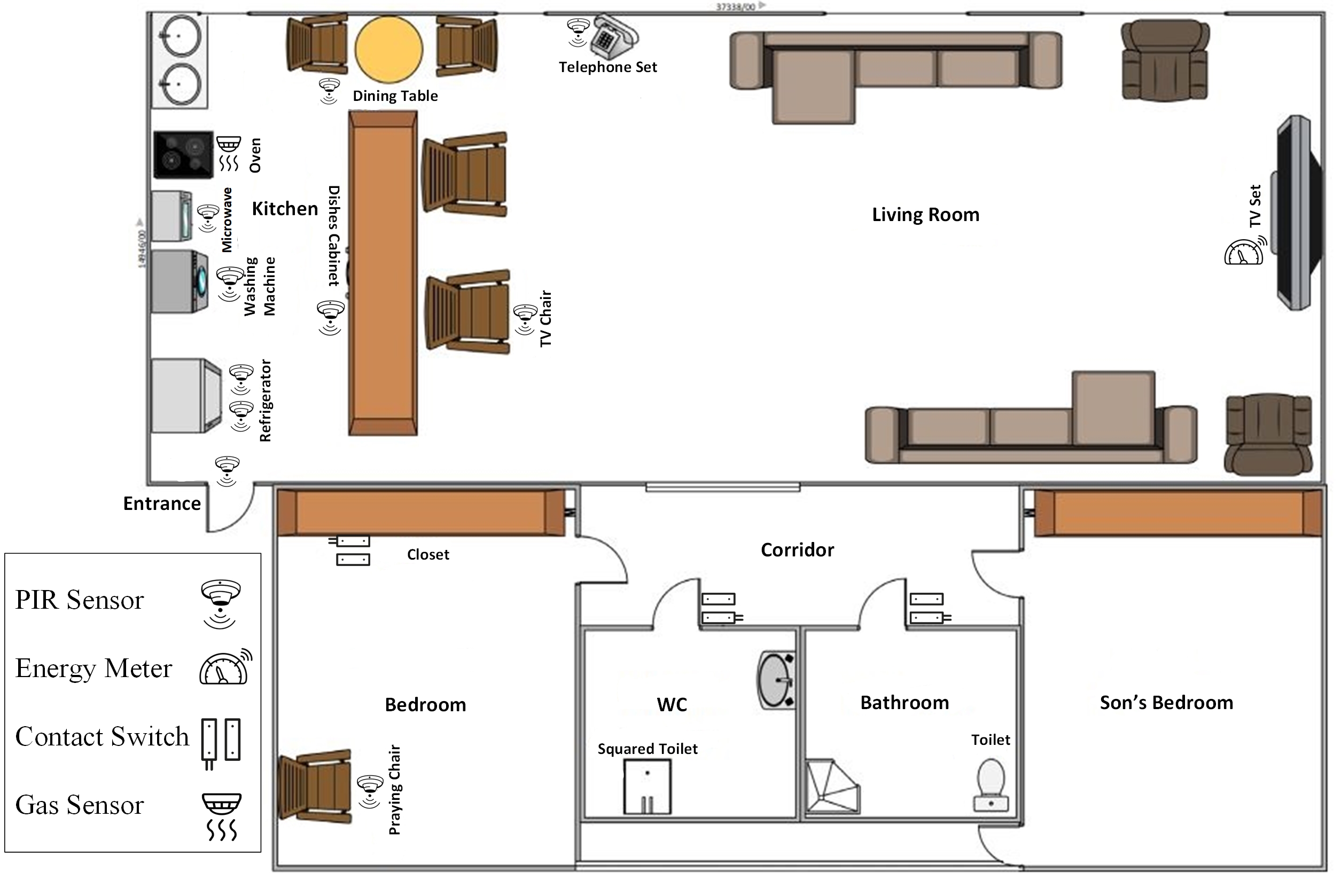}
        \caption{}
        \label{fig1-homeplan}
    \end{subfigure}
    \begin{subfigure}[h]{0.3\textwidth}
        \centering
        \includegraphics[width=\textwidth]{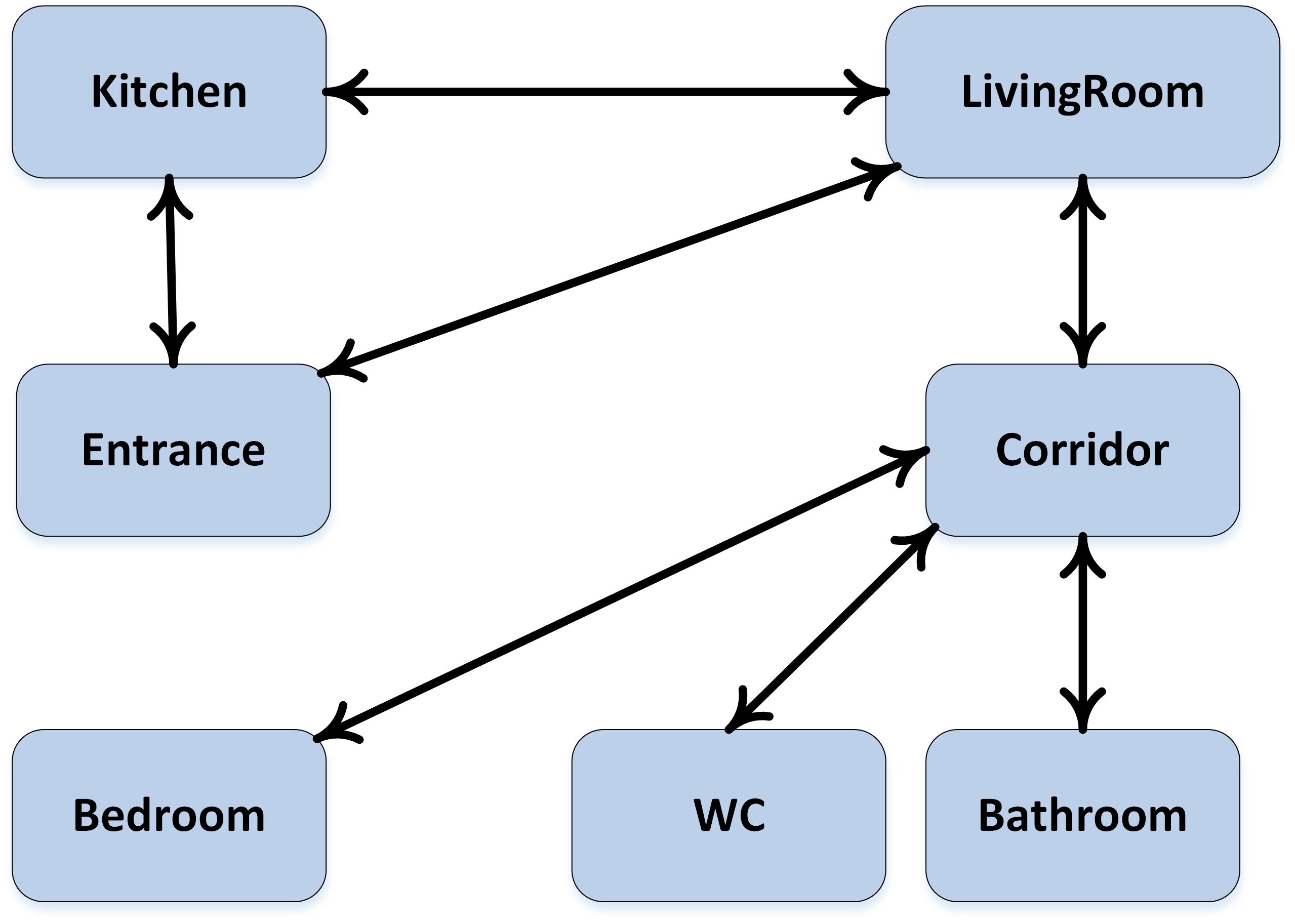}
        \caption{}
        \label{fig2-Possibletransitions}
    \end{subfigure}
    \caption{(a) The house floor plan, location and types of ambient sensors (b) Possible transitions between various areas of the house.}
\end{figure}

The FR-Dataset is collected by using \textit{15} ambient sensors in total, which were installed in \textit{six} different areas of a house. The sensor locations and their types are shown in Figure~\ref{fig1-homeplan}.

The pre-processing step of our study uses a set of simple rules over the records generated by the PIR and contact sensors readings to discover the subject’s presence in \textit{seven} different areas of the house and assigns the following labels to the discovered events at each place: Kitchen, LivingRoom, Bedroom, Corridor, WC, Bathroom and Entrance\footnote{The specific location names of the areas in the FR-dataset are written with capital letters, like Kitchen or LivingRoom. However, small letters are used to refer to places in general.}.
The duration of each event is also calculated by considering the time difference between the first sensor reading in that location and the time of the first reading by sensors in another area. 


Figure~\ref{fig1-homeplan} also illustrates the walls, doorways and furniture of the house, which dictate the possible transitions of the subject between adjacent areas inside the house.
Indeed, using the house’s floor plan, we can easily determine the valid transitions between different areas of the house (Figure~\ref{fig2-Possibletransitions}).
Later, these possible transitions can be used to check if the sequence of captured events is reasonable and to verify the correctness of observed events.

\subsubsection{Dataset errors and their sources}
\label{Dataset-errors}
Error correction methods are helpful for cleaning data from unavoidable errors caused by IoT systems intrinsic characteristics and improving the data quality and subsequently the analysis accuracy.
But, cleaning errors from a log is only possible by discovering their origins.
Indeed, to discriminate actual events from noisy captured events, we should first understand why the errors are generated and then remove the events, that were possibly added to the dataset owing to those error sources.

As mentioned, behaviours are composed of sequences of routines and series of activities at the lower level~\cite{di2021human}.
Regarding the FR-dataset, in fact, the purpose of the system is to detect specific activity times and durations to map them into behaviour patterns.
Hence, the sensors are installed in different areas of the house to highlight the time of using some home’s utilities or discovering some specific activities, such as cooking, watching TV, hygiene activities, or doing home chores.

In this study, we use the available sensor readings from the FR-dataset, corresponding sensor locations and their triggering timestamps to extract the location information and create a log which we refer to as the \textit{location event log}.
However, due to the lack of sensors in doorways, it is not possible to precisely determine the exact time of resident's transitions between the areas.
Also, there are many undiscoverable areas on the home floor plan, which are not covered by any sensors, like the Corridor, son's bedroom and most part of LivingRoom (out of the sensing range of the two existing PIRs).
Thus, the applied location discovery algorithm may produce some unintended errors in the identified locations, the timestamps and sequence of events and the transitions between places.
This study's objective is to address these issues and make the necessary corrections to provide a reasonable sequence of events.

\begin{figure}[t]
\centering
\includegraphics[width=0.7\textwidth]{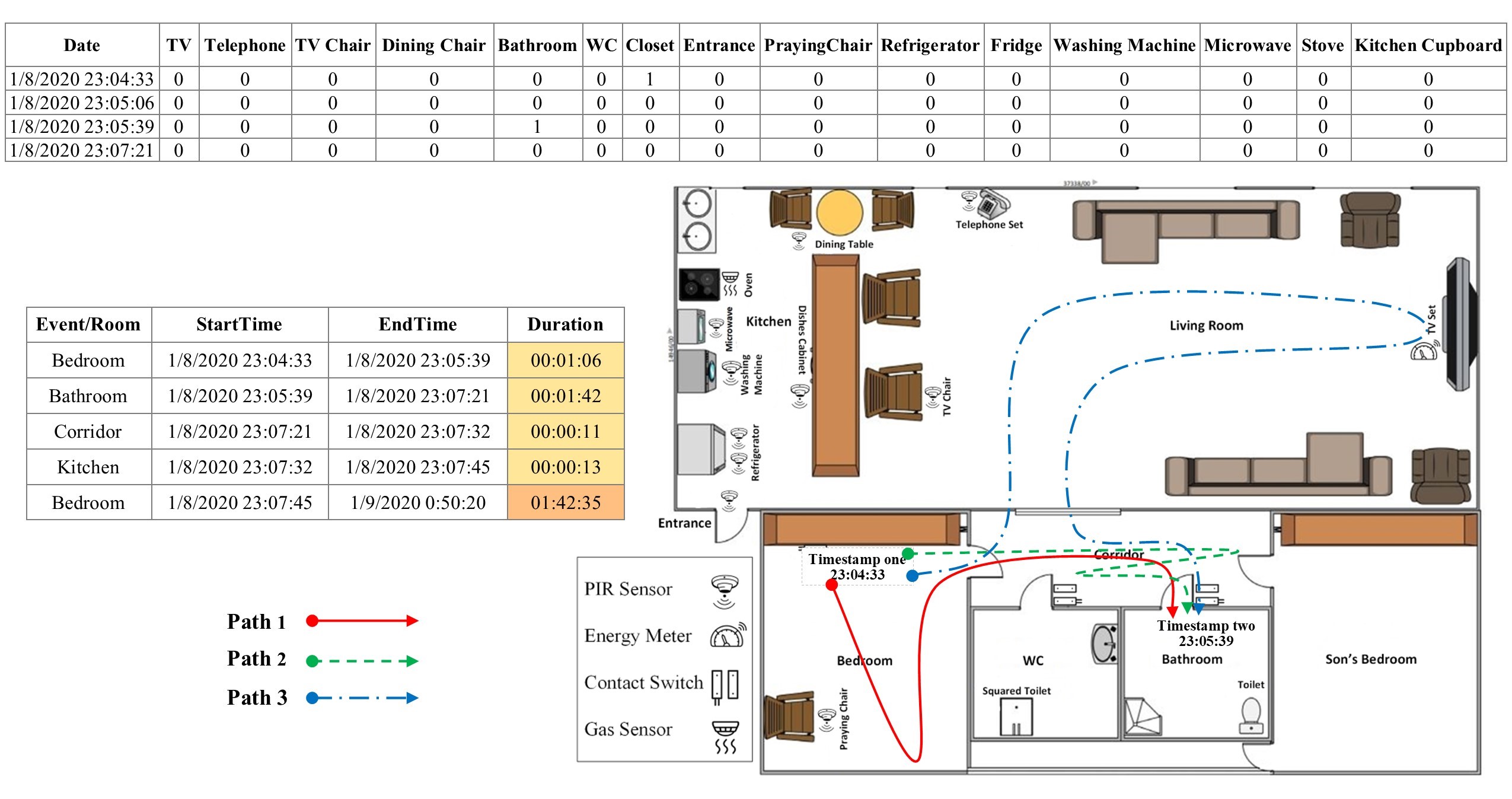}
\caption{An example piece of actual log and the corresponding detected events.}
\label{fig3-example}
\end{figure}

To demonstrate what kind of errors are expected to be generated by the pre-processing algorithm, consider the example illustrated in Figure~\ref{fig3-example}.
It shows a part of the sensor log that contains all sensor reports on the mentioned timestamps, the extracted corresponding location event log, and three sample paths drawn on the map. 
Each \textit{one} value in the sensor log shows a data record (\textit{e.g.} an actual event or activity) captured by a sensor, and the timestamp of that row indicates the occurrence time of the sensor measurement, which in fact expresses the exact time of the subject's movement within the sensing area or her interaction with a utility in the home.

Assuming the data log and discovered locations in the event log in Figure~\ref{fig3-example}, the first timestamp (1/8/2020 23:04:33) shows the time of the subject’s presence in the Bedroom and the second timestamp (1/8/2020 23:05:39) shows the time when the subject opened the Bathroom door and entered the Bathroom.
The Corridor is one of the undiscoverable areas of the house, hence, given the two timestamps mentioned in this particular case, it's difficult to precisely determine when the subject left the Bedroom and entered the Corridor.
The subject could spend more time in the Bedroom and finally go to the Bathroom by passing the Corridor (following Path 1), go to the Corridor and spend some time there (Path 2), or even it is possible for the subject to leave the Bedroom and visit other undiscoverable spaces such as the LivingRoom and then enter the Corridor again and Bathroom at timestamp two (drawn as Path 3).

The specific times of entering and leaving certain locations like Bedroom and LivingRoom cannot be determined with certainty, except for the WC and Bathroom which have micro-switch sensors.
As a result, the required location event log must rely on estimations based on previous and subsequent log entries.
Despite efforts made by the pre-processing algorithm to generate accurate predictions based on captured sensor readings, there may be instances where some events are missed or incorrectly identified.

Nevertheless, it is expected that these predictions and undiscoverable areas could lead to a high amount of errors in the captured events.
Subsequently, the missed transitions result in some mismatches between the discovered sequence of events and the possible transitions defined based on the floor plan of the house (Figure~\ref{fig2-Possibletransitions}).
Every pair of consecutive events that indicates an impossible direct transition between two areas of the home is labelled as an error (such as direct transitions between the Bedroom and Kitchen).
The validation of the transitions in the location event log showed that 4269 out of 12494 transitions (34.16\%) are invalid.

According to this preliminary analysis of event log quality, it contains many errors, and we have to try to remove or correct them (or at least part of them) by applying an error correction algorithm. 
In the following section, the proposed and implemented error correction methods are described.
\section{Error Correction methods}
\label{proposed_techniques}

As previously mentioned in Section~\ref{Section2:backgrounds}, when collecting data using IoT devices, errors such as missing events and noises are inevitable and our case study log, \textit{location event log} also had some data quality issues.

In our smart home scenario, we first discover the reason for capturing erroneous events by analysing the intrinsic specifications of the installed sensors, such as sensing range and sampling rate.
To achieve this, the validity of travelled paths and the sequence of visited areas in the house are evaluated to discover the errors and identify the most problematic sensors.

In the next step, we used error correction techniques adjusted for our case study to correct the discovered errors.
We have contributed to using the knowledge about the most frequent and problematic error sources to highlight events with a higher likelihood of being errors, then inspect and correct them in a heuristic and semi-supervised manner.



\subsection{Preliminary PM-based error correction technique~\cite{Fernandez-Llatas2021RTLS}}
\label{preliminary-PM}
The PM-based correction technique~\cite{Fernandez-Llatas2021RTLS} detects all impossible transitions and defines a penalization score for the possible correction options.
The penalization score is a value, which the algorithm calculates for each single correction option to indicate the amount of changes that are needed based on their impact on the actual log.
After measuring the penalization score for all possible correction options, the algorithm selects the correction option with the minimum penalization score and corrects the actual trace accordingly.
Long-duration events are typically encompass multiple sensor records in the sensor log, abstracted in a long-duration event, which means that there are fewer chances for them to be errors.
The penalization score for long-duration events is high, and they should only change if no other correction with less penalty exists.
The PM-based technique corrects the discovered errors by adding or removing transitions to create a valid trace with the minimum penalty score.
Moreover, the PM-based method needs a reference model, which includes valid transitions (such as Figure~\ref{fig2-Possibletransitions}) to inspect the actual traces and correct the errors.   

\begin{figure}[h]
\centering
\includegraphics[width=0.9\textwidth]{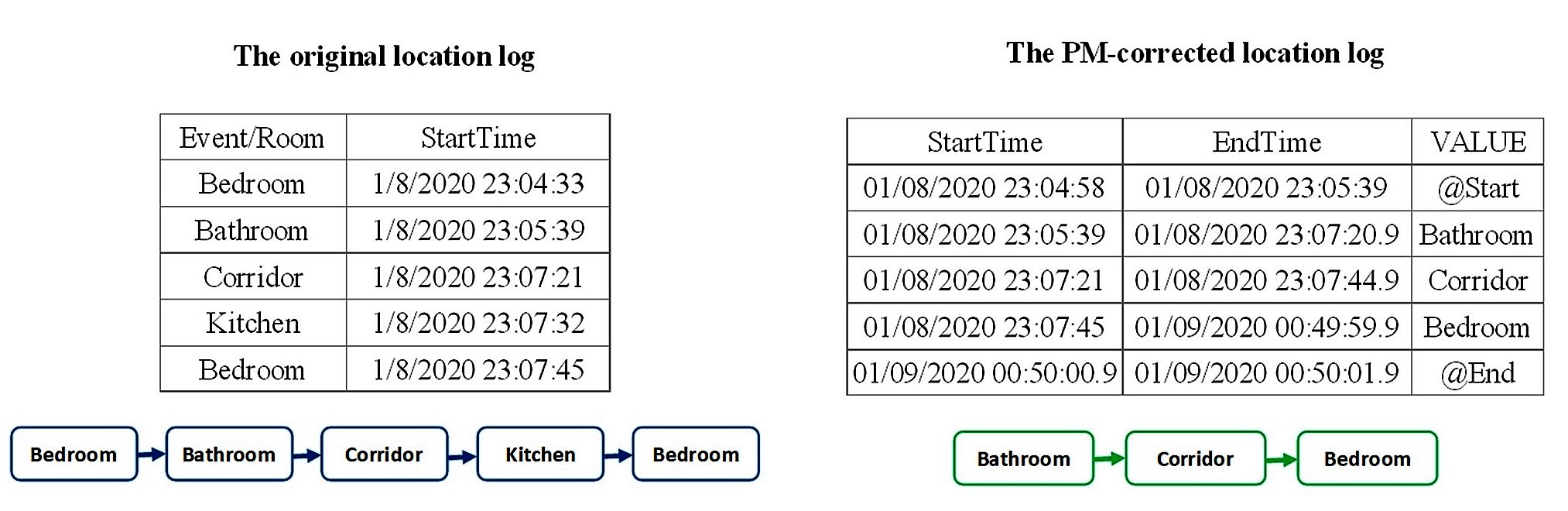}
\caption{The output of preliminary PM-correction method.\label{fig4-output_PM1}}
\end{figure}   
\unskip

\subsection{Rule-based error correction}
\label{Rule_based}
To correct the errors in a log using a rule-based approach in a semi-supervised way, a set of rules should be heuristically defined to highlight possible errors in the entire log for further inspection, and as explained in section~\ref{sec:errorCorrectionsGeneral}, it has been used in multiple studies so far.
Our contribution is to integrate the knowledge of IoT devices, their challenges, and the reasons for capturing errors (as described in Section~\ref{Section2.2-error}) into the rule-determination process.
 
For instance, experts can check the certainty of highlighted location events by counting the number of sensor records that occurred within the time interval of those events and decide to remove them if the events are recognised as possible errors.
Therefore, if accurate rules are defined based on the expected behaviour of the system and applied to the log, this rule-based approach can find unusual and incorrect events, and then clean the dataset from noisy and false detections.

We have contributed to the error correction problem by using rules, which are defined based on recognised weaknesses of IoT systems, as identified by the experts, to correct noise-type erroneous events.
Two types of rules are determined in the proposed rule-based error correction approach:
\begin{itemize}
\item	\textbf{Type 1)} Events with unusual long-duration.
For the places, which we know the subject uses for a short time (WC and Bathroom) or just for transition and reaching other areas (Corridor) a threshold for the maximum duration is used for highlighting potential noises.
   
\item	\textbf{Type 2)} Events with duration less than usual.
A minimum duration threshold can highlight the unusual events of each place for further inspection.
A minimum threshold is used just for the places where we expect the subject to be there for a long time, \textit{i.e.} LivingRoom, Kitchen, Entrance and Bedroom.
\end{itemize}

The final applied rules are summarised in Table~\ref{table:min_Max_Rule_based_thr}.
\begin{table}[H] 
\caption{Determined minimum and maximum threshold for rule-based noise correction.\label{table:min_Max_Rule_based_thr}}
\begin{tabular}{cccc}
\toprule
\textbf{Location}	& \textbf{Min Threshold}	& \textbf{Max Threshold}  & \textbf{Applied rule}\\
\midrule
Bathroom		& -			& 00:08:29	 		& Average + 2 ×StdDev\\
Bedroom			& 00:00:11			& 05:27:48			& 2.5\% rule on both sides\\
Corridor			& -			& 00:08:29			& Average + 2 ×StdDev\\
Entrance			& 00:01:47			& 04:38:10			& 2.5\% rule on both sides\\
Kitchen			& 00:00:06			& 00:52:10			& 2.5\% rule on both sides\\
LivingRoom			& 00:00:30			& 00:59:21			& 2.5\% rule on both sides\\
WC			& -			& 00:04:47			& Average + 2 ×StdDev\\

\bottomrule
\end{tabular}
\end{table}
\section{Results and Discussion}
\label{Section4:results-discussion}
We evaluated the effectiveness of the adapted existing methods, for error detection and correction by applying them to the FR-dataset.
Our primary objective was to observe how these techniques can detect and fix errors present in the log, which contains data collected by an IoT system in a real-life setting.

\subsection{Results of Rule-based error correction}
\label{results-Rule_based}
Table~\ref{table:results_Rule_based}, lists the results of implementing the rule-based error correction approach, including the number of events before applying the correction method, the number of events after correction, the total number of changes (total altered records, \textit{e.g.} added, removed or fused events) and the percentage of changes in relative to total events in each location.
As can be noticed from Table~\ref{table:results_Rule_based}, only 3.218\% of changes were made to the entire log to correct the traces based on the determined rules and experts' knowledge.
This percentage is lower than our predetermined correction target, indicating that the log was only slightly altered after correction.

\begin{table}[h] 
\caption{The outcome of applying the rule-based error correction technique on FR-dataset\label{table:results_Rule_based}}
\setlength{\tabcolsep}{5pt} 
\begingroup
\begin{tabular}{ccccc}
\toprule
Locations	& Number of Events	& After Correction Events  & Total Corrections (\%)\\
\midrule
Bathroom		& 1655		& 1653		& 10 (0.605\%) \\
Bedroom		& 1436		& 1369		& 86 (6.282\%) \\
Corridor		& 2371		& 2375		& 133 (5.600\%) \\
Entrance		& 242		& 233		& 17 (7.296\%) \\
Kitchen		& 3279		& 3179		& 158 (4.970\%) \\
LivingRoom		& 2795		& 2842		& 92 (3.237\%) \\
WC		& 716		& 716		& 4 (0.559\%) \\
\midrule
Total		& 12494		& 12367		& 398 (3.218\%) \\

\bottomrule
\end{tabular}
\endgroup
\end{table}

\subsection{Evaluation of error correction techniques}
\label{evaluation}
Finally, to evaluate the adapted error correction techniques, we have measured log change metrics.

\begin{table}[h]
\caption{A summary of log change metrics for correction methods}
\label{tab:evaluation_results}
\begin{tabular}{p{0.2\textwidth}p{0.09\textwidth}p{0.08\textwidth}p{0.08\textwidth}p{0.12\textwidth}p{0.12\textwidth}p{0.115\textwidth}}
\toprule
 & \multicolumn{3}{c}{\textbf{Number of Changes}}  & \multicolumn{3}{c}{\textbf{Duration of Changes}} \\ \cmidrule{2-4} \cmidrule{5-7}
 & \multicolumn{2}{c}{\textbf{Daily}} & \textbf{Total} & \multicolumn{2}{c}{\textbf{Daily}} & \textbf{Total} \\ \cmidrule{2-3} \cmidrule{5-6}
\textbf{Error Corrections}                       & \textbf{Average} & \textbf{Max} & \textbf{}      & \textbf{Average} & \textbf{Max} & \textbf{}      \\ 
\midrule
\textbf{Rule-based on ORL} & 2 & 9 & 256 (2.07\%) & 00:02:18 (0.160\%) & 00:27:35 (1.916\%) & 5:35:41 (0.160\%) \\
\midrule
\bottomrule
\end{tabular}
\end{table}

Based on the evaluation results, it is evident that the semi-supervised rule-based algorithm is more effective in managing noises.
This is because experts possess a better comprehension of typical behaviours, which significantly reduces the probability of inaccurate corrections.

\section{Conclusions}
\label{Section5:Conclusions}

If the data quality of IoT devices is found to be low, error correction algorithms can be employed to correct the errors.
This article highlights that by identifying the main reasons for capturing errors during the data collection, the error correction methods can be adapted in a way to address the errors better.
A rule-based method with specifically determined rules to address the IoT system limitations is used to correct noises with higher accuracy.


\footnotesize
\bibliographystyle{ieeetr}
\bibliography{refs.bib}


\end{document}